\documentclass[aps,pre,reprint, amsmath, amssymb,superscriptaddress]{revtex4-1}

\usepackage{bm}
\newcommand{\beq}{\begin{equation}}
\newcommand{\eeq}{\end{equation}}

\usepackage{graphicx,tikz,placeins}
\usepackage{mathrsfs}
\usepackage{amsmath,amssymb,amsfonts}
\usepackage{color}
\usepackage{mathrsfs}
\usepackage{float}
\usepackage{subcaption}
\usepackage{times,txfonts}
\usepackage{nicefrac}
\usepackage[colorlinks=true,linkcolor=blue,urlcolor=blue,citecolor=blue,pdfusetitle]{hyperref}
\usepackage{soul}
\usepackage[normalem]{ulem}

\graphicspath{{Figures/}}

\begin{document}

\title{Universal splitting of phase transitions and performance optimization in driven collective systems}
\author{Gustavo A. L. For\~ao}
\affiliation{Universidade de São Paulo,
Instituto de Física,
Rua do Matão, 1371, 05508-090
São Paulo, SP, Brazil}

\author{Jonas Berx}
\affiliation{Niels Bohr International Academy, Niels Bohr Institute,
University of Copenhagen, Blegdamsvej 17, 2100 Copenhagen, Denmark}

\author{Tan Van Vu}
\affiliation{Center for Gravitational Physics and Quantum Information, Yukawa Institute for Theoretical Physics, Kyoto University, Kitashirakawa Oiwakecho, Sakyo-ku, Kyoto 606-8502, Japan}

\author{Carlos E. Fiore}
\affiliation{Universidade de São Paulo,
Instituto de Física,
Rua do Matão, 1371, 05508-090
São Paulo, SP, Brazil}

\date{\today}

\begin{abstract}
Spontaneous symmetry breaking is a hallmark of equilibrium systems, typically characterized by a single critical point separating ordered and disordered phases. Recently, a novel class of non-equilibrium phase transitions was uncovered [Phys. Rev. Res. {\bf 7}, L032049 (2025)], showing that the combined effects of simultaneous contact with thermal baths at different temperatures and external driving forces can split the conventional order-disorder transition into two distinct critical points, determined by which ordered state initially dominates. We show the robustness of this phenomenon by extending a minimal interacting-spin model from the idealized case of simultaneous bath coupling to a finite-time coupling protocol. In particular, we introduce two protocols in which the system interacts with a single bath at a time: a \emph{stochastic} protocol, where the system randomly switches between the baths at different temperatures, and a \emph{deterministic} protocol where the coupling alternates periodically. Our analysis reveals two key results: (i) the splitting of phase transitions persists across all coupling schemes---simultaneous, stochastic, and deterministic---and (ii) different optimizations of power and efficiency in a collectively operating heat engine reveal that both the stochastic and deterministic protocols exhibit superior global performance at intermediate switching rates and periods when compared to simultaneous coupling. The global trade-off between power and efficiency is described by an expression solely depending on the temperatures of thermal reservoirs as the efficiency approaches to the ideal limit.
\end{abstract}

\maketitle

{\it Introduction}---The dynamics and thermodynamics of collectively operating many-body complex systems have recently garnered significant interest due to the presence of novel features as compared to single-unit systems, such as phase transitions, ergodicity breaking, chaos, and the possibility of designing thermodynamic heat engines with superior performance compared to ones composed from single units~\cite{Imparato_2015,herpich2,herpich,filho2023powerful,mamede2021obtaining,lynn2021broken}. Many of these phenomena stem from heterogeneous interactions between the system's constituents. The interplay between these interactions and other aspects, such as temperature or external driving, can give rise to multi-critical behavior. Although these mechanisms are not yet fully understood in the realm of non-equilibrium statistical mechanics, they are now actively explored in both classical~\cite{PhysRevLett.131.017102,gatien,gatien2,herpich,mamede2023,filho2023powerful} and quantum systems~\cite{friedemann2018quantum,rolandi,Hagman2025}.

Recently, a novel class of non-equilibrium phase transitions was introduced~\cite{foraohzl3-hjnl}, arising in collective systems \emph{simultaneously} exposed to multiple heat baths, as well as driving forces. These conditions give rise to order-disorder transitions that differ strikingly from those found in conventional equilibrium and non-equilibrium settings. A key aspect is that the transition point depends on the initially ordered phase, and the resulting transitions can belong to different classes: the transition emerging from the ``down''-spin phase is continuous, whereas the one from the ``up''-spin phase is discontinuous. Together with modified critical exponents, this feature shows that a single system can exhibit fundamentally different types of phase transitions under identical control parameters. 

Beyond their distinct phase behaviors, such collectively operating systems also exhibit a rich variety of novel thermodynamic traits.
Within this broader thermodynamic framework, collective heat engines emerge as systems whose operation can outperform their non-interacting counterparts, displaying reduced dissipated work~\cite{rolandi,Meibohm24}, higher efficiency or improved power~\cite{gatien,filho2023powerful,mamede2023,mamede2025collective}. The presence of, e.g., non-conservative driving forces and simultaneous contact between different heat baths also leads to direct implications for thermodynamic performance, giving rise to two major differences compared to single and other collective heat engines: (i) only the ``down''-spin ordered phase can operate as a heat engine or heat pump, exhibiting lower entropy production and smaller power fluctuations than the ``up''-spin ordered phase, and (ii) under certain conditions such heat engines are able to be designed to operate close to the regime of optimal efficiency at maximum power~\cite{foraohzl3-hjnl}. 

In this work, we go beyond the idealized description of simultaneous coupling by introducing non-simultaneous contact with the thermal reservoirs. This places the engine in a finite-time operational regime, making the interplay between dissipation, emergent order, and cycle duration central to its performance. In this framework, the relevant questions are no longer limited to phase behavior, but also concern how collective systems navigate power-efficiency trade-offs. To this end, we propose and analyze two contrasting protocols: a \textit{stochastic} coupling, in which the system randomly switches between reservoirs at a rate $d$ and a \textit{deterministic} one, where the coupling alternates periodically every half-cycle of fixed duration $\tau/2$.



{\it Setup and non-conservative driving}---We consider a minimal model composed of \( N \) interacting units, each represented by a spin variable \( s_j \in \{-1, 0, 1\} \). The system is alternately coupled to thermal baths at temperatures $T_\nu=(k_B\beta_\nu)^{-1}$, with $\nu=1,2$. A microscopic configuration of the spins is defined by the vector \( S  \equiv (s_1, \ldots, s_N) \) and the system energy $E$ takes the form
\begin{equation}
E(S) = \epsilon \sum_{(i,j)} s_i s_j + \Delta \sum_{i=1}^N s_i^2,
\label{eqq}
\end{equation}
where \( \epsilon < 0 \) quantifies the interaction energy between two nearest neighbors, denoted by the pair $(i,j)$, and each unit contributes an individual energy \( 0 \) or \( \Delta \). The interactions among constituents give rise to collective effects, leading to phase transitions as the parameters \( \epsilon \) or \( \Delta \) are varied. For large \( |\epsilon| \),
the system becomes confined to two ordered phases, \( A \) and \( B \), characterized respectively by a predominance of spins \( -1 \) and \( +1 \). Defining the order parameter \( m = -\langle \sum_{i=1}^{N}s_i \rangle/N \), where $\langle...\rangle$
denotes the ensemble average, we have \( m > 0 \) in phase \( A \), \( m < 0 \) in phase \( B \), and \( m = 0 \) for disordered configurations.

Our analysis focuses on all-to-all interactions, by taking the summation in Eq.~(\ref{eqq}) over all possible pairs of spins and replacing $\epsilon \rightarrow \epsilon/N$ such that the system configuration is characterized by the total number of spins $N_{s}$ in each state $s\in\{0,\pm 1\}$. The system energy in Eq.~(\ref{eqq}) then becomes
\begin{equation}
E(S) \rightarrow \frac{\epsilon}{2N}\left\{ \sum_{s=\pm1}N_{s}\left(N_{s}-1\right) - 2N_{+1}N_{-1} \right\} +\Delta\sum_{s=\pm1}N_{s}\,.
\label{alltoall}
\end{equation}
We assume that the dynamics exclusively consists of single spin flips; that is, at any given time, only one spin transitions from a state $s' \in \{0,\pm1\}$ to $s \in \{0,\pm1\}$, with $s \neq s'$. This corresponds to updating the occupation numbers as $N_{s'} \rightarrow N_{s'} -1$ and $N_{s} \rightarrow N_{s} + 1$. The transition rates are in Kramers' form, $\omega^{(\nu)}_{ss'} = \Gamma \exp\!\left[-\frac{\beta_\nu}{2}Q_{ss'}^{(\nu)}\right]$, where $\Gamma$ sets the (global) characteristic timescale of the process, and \( Q_{ss'}^{(\nu)} \) denotes the amount of heat absorbed from the $\nu$-th thermal bath during the transition \( s' \to s \), given by
\begin{equation}
Q_{ss'}^{(\nu)} = \Delta E_{ss'} - W_{ss'}^{(\nu)}\,.
\end{equation}
Here, \( \Delta E_{ss'} \) is the change in internal energy between configurations \( s \) and \( s' \), and $W_{ss'}^{(\nu)}$ represents the amount of work performed on the system.

\begin{figure}
    \centering
    \includegraphics[width=0.95\linewidth]{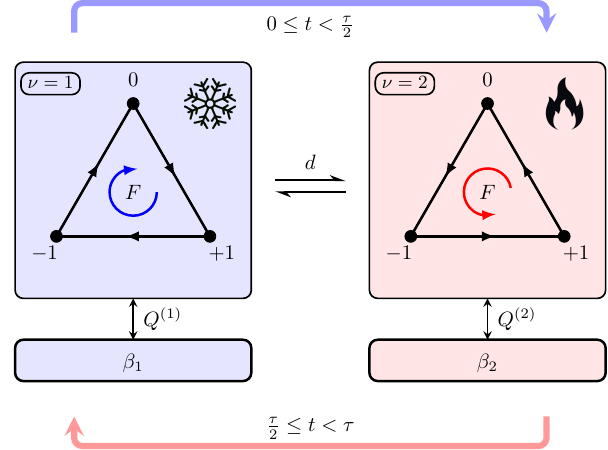}
    \caption{Schematics and dynamics of the collective engine. The system is composed of an ensemble of all-to-all interacting units, each one described by the single-spin variable $s\in \{0,\pm1\}$ (triangle vertices). Left and right panels show the set of transitions $s' \rightarrow s$ (black arrows) which are favored by the biased driving $F$ according to whether the system is placed in contact with the cold ($\nu=1$; blue) or hot ($\nu=2$; red) thermal reservoirs, at inverse temperatures $\beta_\nu$. \emph{Stochastic} switching between thermal reservoirs, given the system is at the state $s$, occurs with rate $d$ (black, double arrows); \emph{deterministic} switching occurs periodically with period $\tau$ (outer arrows; colored according to their contact with the respective reservoir).
    }
    \label{fig0}
\end{figure}

In the thermodynamic limit ($ N\rightarrow \infty$), the fraction of spins in state $s$, given by $N_{s}/N$, is identified with the state occupation probability $p_{s}^{(\nu)}$. We consider the representative local configurations illustrated in Fig.~\ref{fig0}, where the work performed on the system is determined by a non-conservative driving force, i.e.,
\begin{equation}
W_{ss'}^{(\nu)} = (-1)^{\nu-1} d_{ss'} F\,,
\end{equation}
where \( F \) denotes the driving strength, and \( d_{ss'} = +1\) if the transition from \(s' \to s \) follows the clockwise direction along the cyclic sequence \( -1 \rightarrow 0 \rightarrow +1 \rightarrow -1 \). Conversely, \( d_{ss'} = -1\) if the transitions occur in the counterclockwise direction. Thus, for \( F >0 \), the cold bath (\( \nu = 1 \)) promotes clockwise spin flips, whereas the hot bath (\( \nu = 2 \)) drives counterclockwise ones (see Fig.~\ref{fig0}). Henceforth, we will simplify our index notation by denoting the states $s =\pm1$ as $s = \pm$ instead, omitting the ``1''.
As a result of the driving, the contributions to the  heat take the form
\begin{equation}
    \label{eq:Qs}
    \begin{split}
        Q_{0-}^{(\nu)} &= \epsilon \big(p^{(\nu)}_+ - p^{(\nu)}_-\big) - \Delta - (-1)^{\nu-1}F\,,\\
        Q_{+0}^{(\nu)} &= \epsilon \big(p^{(\nu)}_+ - p^{(\nu)}_-\big) + \Delta - (-1)^{\nu-1}F\,,\\
        Q_{-+}^{(\nu)} &= 2\epsilon \big(p^{(\nu)}_- - p^{(\nu)}_+\big) - (-1)^{\nu-1}F\,,
    \end{split}
\end{equation}
corresponding, respectively, to the transitions \( -1 \to 0 \), \( 0 \to +1 \), and \( +1 \to -1 \), each induced by the $\nu$-th bath. 

{\it Simultaneous contact between thermal baths---}The existence of distinct phase transitions that depend on which ordered phase initially dominates was recently demonstrated for systems in simultaneous contact with two thermal baths~\cite{foraohzl3-hjnl}. It contrasts sharply with standard order–disorder transitions, where the order parameter typically vanishes as $ |m|\sim\sqrt{a(\epsilon_c^{A} - \epsilon)}$, with \( a > 0 \), when $\epsilon$ approaches the critical point $\epsilon_c^A$. For the system considered here, with $\Delta = 0,F \neq 0$ and $\beta_1\neq \beta_2$, the two ordered phases behave fundamentally differently: phase \( B \) exhibits a discontinuous transition at $\epsilon_c^{B}$ while phase \( A \) follows a continuous scaling \( m \sim a(\epsilon_c^{A} - \epsilon) \), where the critical point $\epsilon_c^A$ is given by~\cite{foraohzl3-hjnl}

\begin{equation}
\epsilon_c^{A} = -\frac{e^{\frac{1}{2} F (\beta_1 - \beta_2)} + e^{\frac{1}{2} F (\beta_2 - \beta_1)} + e^{\frac{1}{2} F (\beta_1 + \beta_2)} + e^{\beta_1 F} + e^{\beta_2 F} + 1}{\left(e^{\frac{\beta_1 F}{2}} + e^{\frac{\beta_2 F}{2}}\right) \left(\beta_1 \cosh \left(\frac{\beta_1 F}{2}\right) + \beta_2 \cosh \left(\frac{\beta_2 F}{2}\right)\right)}\,,
\label{eq2}
\end{equation}
with $\epsilon_c^{B}\neq \epsilon_c^{A}$.
Both  transitions can be understood by assuming the order parameter satisfies the expansion $a(\epsilon - \epsilon_c^{A}) m + b m^2 + c m^3 + \ldots = 0$, where $a,b,c>0$ are coefficients that generally depend in a non-trivial way on system parameters, but can be evaluated numerically. The coefficient $b$, however, admits the factorized form  $b = f \left( F(\beta_1 - \beta_2)\right) g(\beta_1, \beta_2, F)$, with $f(0)=0$ since $b=0$ whenever $F=0$ or $\beta_1=\beta_2$. While both solutions $m>0$ and $m<0$ are stable for $\epsilon<\epsilon_c^{A}$, the $m>0$ solution smoothly vanishes as $\epsilon\rightarrow\epsilon_c^{A}$, consistent 
with a continuous phase transition and \( m \sim a(\epsilon_c^{A} - \epsilon) \). To describe the discontinuous transition, it is convenient to first rewrite the above expansion in the following form:  $m^3 + \mu m^2 - \lambda m = 0$, with $\lambda = -\frac{a}{c}(\epsilon-\epsilon_c^A)$ and $\mu = b/c$, which is the canonical form of a perturbed pitchfork bifurcation~\cite{Landau1996,Golubitsky1984}. In this geometry, the branch of solutions $m<0$ is unstable for $\epsilon>\epsilon_c^A$, bends back and becomes a stable branch. Since the system only realizes stable states, the transition instead appears as an abrupt jump from $m=0$ to a finite $m<0$ at the value $\epsilon_c^B > \epsilon_c^A$. Similar results can be computed for $\Delta\neq 0$, although in this case both phase transitions become discontinuous as $\Delta$ increases~\cite{foraohzl3-hjnl}.

We now extend this framework by introducing finite-time reservoir switching, moving beyond the idealized simultaneous-coupling scenario. We consider both \textit{stochastic} switching, where the system probabilistically changes contact at a constant transition rate, and \textit{deterministic} switching, where it alternately couples to the hot and cold reservoirs for fixed durations.

{\it Stochastic switching---}The stochastic description can be interpreted as a ``two-box'' model~\cite{danielPhysRevResearch.2.043257,busiello2021dissipation,mamede2025collective}, in which each unit interacts with one heat bath at a time, while transitions between baths are instantaneous and occur without changing the unit's spin state. The time evolution of probability ${p}^{(\nu)}_{s}(t)$ reads
\begin{equation}\label{seq_stoc}
\dot{p}^{(\nu)}_{s}(t)=\sum_{s' \neq s}J^{(\nu)}_{ss'}(t)+\sum_{\nu' \neq \nu}\mathcal{K}^{(s)}_{\nu \nu'}(t),\end{equation} where the first term on the right-hand side accounts for spin-flip dynamics at fixed $\nu$, i.e.,
\begin{equation} 
J^{(\nu)}_{ss'}(t)=   \omega^{(\nu)}_{ss'} p^{(\nu)}_{s'}(t) - \omega^{(\nu)}_{s's} p^{(\nu)}_s(t)\,.
\label{mee2}
\end{equation}  
Similarly, the second term in Eq.~\eqref{seq_stoc} accounts for exchanges between thermal baths at fixed spin state $s$ and is given by
\begin{equation}
\mathcal{K}^{(s)}_{\nu \nu'}(t)=\Omega^{(s)}_{\nu \nu'}  p^{(\nu')}_{s}(t)-\Omega^{(s)}_{\nu'\nu}  p^{(\nu)}_s(t)\,,
\end{equation}
with $\Omega^{(s)}_{\nu \nu'}$ the switching rate from reservoir $\nu'$ to $\nu$ at fixed state $s$. For the two reservoirs $\nu=1,2$ under consideration, the entropy production rate can be expressed in a compact current-affinity form as
\begin{equation}
    \dot{\sigma}(t) = 
    \sum_{\nu}\sum_{s<s'} J_{ss'}^{(\nu)}(t) X^{(\nu)}_{ss'}(t)
    + \sum_{s} \mathcal{K}_{21}^{(s)}(t) Y_{21}^{(s)}(t),
    \label{eps_Current}
\end{equation}
where $X^{(\nu)}_{ss'}(t) = \ln \!\left[ \omega^{(\nu)}_{ss'} p^{(\nu)}_{s'}(t) / \omega^{(\nu)}_{s's} p^{(\nu)}_{s}(t) \right]$ and $Y_{21}^{(s)}(t) = \ln \!\left[ \Omega^{(s)}_{21} p^{(1)}_{s}(t) / \Omega^{(s)}_{12} p^{(2)}_{s}(t) \right]$ are the corresponding thermodynamic affinities and the sum runs over all distinct spin transitions ($s<s'$).   
Assuming symmetric switching rates, $\Omega^{(s)}_{12} = \Omega^{(s)}_{21} = d$ for $s \in \{0, \pm\}$, two limiting cases can be identified: for $d \to 0$, the system equilibrates with a single bath, whereas for $d \to \infty$, it recovers the regime of simultaneous contact with both reservoirs, which is equivalent to replacing $\epsilon \to 2\epsilon$ in the transition rates $\omega^{(\nu)}_{ss'}$ (see Eq.~\eqref{eq:Qs}).

In the non-equilibrium steady state (NESS), defined by time-independent probabilities $p^{(\nu)}_0$ and $p^{(\nu)}_{\pm}$, the entropy production takes the Clausius-like form $\dot{\sigma} = -\sum_\nu \beta_\nu  \dot{Q}_\nu$ where $\dot{Q}_\nu  = \sum_{s<s'} Q^{(\nu)}_{ss'} J^{(\nu)}_{ss'}$ 
with the currents $J^{(\nu)}_{ss'}$ given by Eq.~(\ref{mee2}) evaluated at steady state. In the present case, the heat fluxes assume the explicit form $ \dot{Q}_\nu  =Q^{(\nu)}_{-0} J^{(\nu)}_{-0} + Q^{(\nu)}_{-+} J^{(\nu)}_{-+} + Q^{(\nu)}_{0+} J^{(\nu)}_{0+}$.
The system performance is quantified via mean output power and efficiency, given by
\begin{equation}
  \mathcal{P} = -\sum_\nu \sum_{s<s'} W^{(\nu)}_{ss'} J^{(\nu)}_{ss'}\,,\qquad {\rm and}\qquad\eta=\frac{ {\cal P}}{ {\dot Q_2}}.
 \label{powers}
\end{equation}
The power is related to   ${\dot Q}_\nu$'s through the first law of thermodynamics, $ -\mathcal{P}  + \dot{Q}_1  +  \dot{Q}_2  = 0$. Henceforth, we restrict our analysis to the heat-engine regime, where the efficiency $\eta$ satisfies the bound $0\le\eta\le \eta_C$, with the Carnot efficiency given by $\eta_C=1-\beta_2/\beta_1$. 

The phase transition is characterized via the order parameter $m=m_1+m_2$, where $m_\nu=p^{(\nu)}_{-}-p^{(\nu)}_{+}$. For $-\beta_\nu \epsilon\gg1$, one expects $m\approx 1$ for phase $A$ and $m\approx -1$ for phase $B$. In contrast, in the disordered phase, one expects $m=0$, since $p^{(\nu)}_{-}=p^{(\nu)}_{+}=1/6$.

{\it Deterministic switching---}In the deterministic description, the system is similarly coupled to only one heat bath at a time, but each contact lasts for a fixed interval: $(\nu-1)\tau/2 \le t < \nu\tau/2$. The switching between baths occurs instantaneously at $t=\tau/2$ (from cold to hot) and $t=\tau$ (from hot to cold). This type of dynamics, often referred to as a sequential or collisional, has been investigated in several contexts, including two-state~\cite{rosas1,rosas2,harunari2020maximal} and three-state~\cite{felipe} Markovian systems, as well as for Brownian particles~\cite{noa2020thermodynamics,noa2021efficient,PhysRevE.110.054125}. Here, we extend this approach to a collective system to explore how the deterministic switching time influences both the phase transitions and the performance of the heat engine, and compare with stochastic switching protocols.  

In the deterministic case, the system dynamics is governed by the master equation
\begin{equation}\label{seq_det}
\dot{p}^{(\nu)}_s(t)=\sum_{s' \neq s}J^{(\nu)}_{ss'}(t),
\end{equation}
where $J^{(\nu)}_{ss'}(t)$ is defined in Eq.~(\ref{mee2}). In contrast to the stochastically switching system, deterministic switching between the two heat baths is imposed through the boundary conditions $p^{(1)}_s(\tau/2)=p^{(2)}_s(\tau/2)$ and $p^{(1)}_s(0)=p^{(2)}_s(\tau)$, for all states $s$. Unlike the stochastic switching case, where the system reaches a time-independent NESS, the deterministic setup leads to a periodic steady state, characterized by time-periodic state occupations $\{p^{(\nu)}_0(t),p^{(\nu)}_{\pm}(t)\}_{0\le t\le\tau}$. The phase transition is again characterized using an order parameter $m=m_1+m_2$, similar to the stochastic case, but with the following modified definition:
\begin{equation}
    \label{eq:m_deterministic}
    m_\nu=\frac{1}{\tau}\int_{({\nu-1})\tau/2}^{\nu\tau/2}\left[p^{(\nu)}_{-}(t)-p^{(\nu)}_{+}(t)\right]\mathrm{d}t\,.
\end{equation}

\begin{figure*}[tp]
    \centering
    \includegraphics[width=0.85\linewidth]{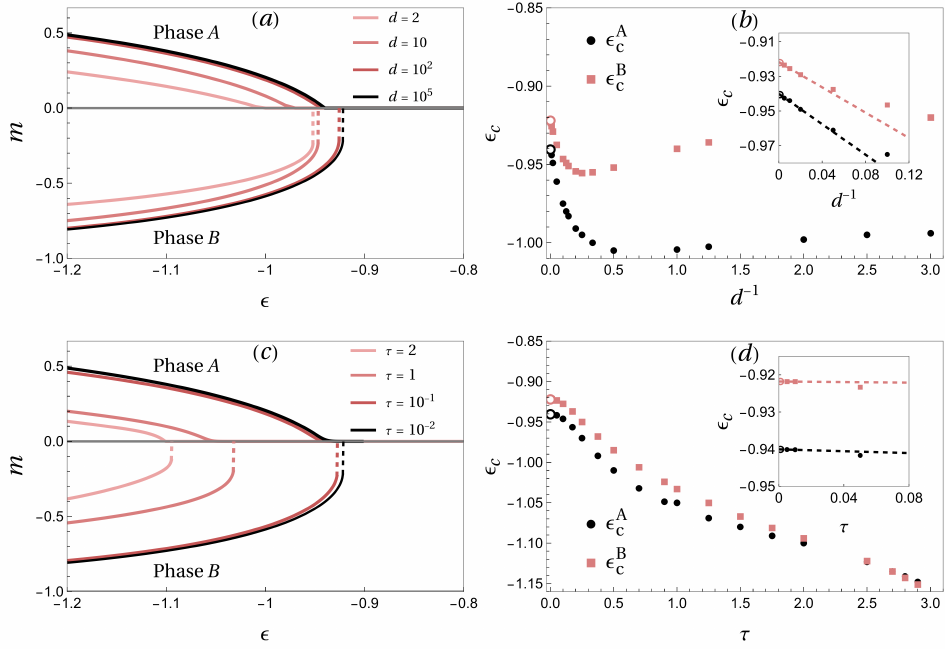}
     \caption{Order parameter $m$ and critical points $\epsilon_c^{A,B}$ as a function of the stochastic switching rate $d$ (a, b) and the deterministic cycle time $\tau$ (c, d). Panels a and c show $m(\epsilon)$ for stochastic and deterministic switching, respectively, illustrating the splitting into phases $A$ and $B$ at distinct critical points $\epsilon_c^{A,B}$ for different values of $d$ and $\tau$. Panels b and d display the corresponding critical points $\epsilon_c^{A}$ (black symbols) and $\epsilon_c^{B}$ (red symbols) as functions of $d^{-1}$ and $\tau$. In both cases, the results converge to the simultaneous-contact limit as $d^{-1}\to0,\,\tau \to 0$. Insets in panels b and d highlight the behavior of the critical points near this limit. Parameters are $F=2,\,\beta_2=1$ and $\beta_1=2$.}
    \label{fig:fig1}
\end{figure*}   

The thermodynamic quantities---entropy production, heat and power---averaged over one complete cycle with period $\tau$ retain the same structure as those in the stochastic switching case, and will henceforth be denoted by overbars. The key difference lies in the time-averaging procedure, which now spans the full cycle period. In particular, both the power output and entropy production can be expressed as 
 \begin{eqnarray}
  {\overline {\cal P}} &=& -\frac{1}{\tau}\sum_{\nu} \sum_{s<s'}\int_{(\nu-1)\tau/2}^{\nu\tau/2} J_{ss'}^{(\nu)}(t)W_{ss'}^{(\nu)} \mathrm{d}t, \label{eq:power_deterministic}\\
    {\overline \sigma}&=&\frac{1}{\tau}\sum_{\nu}\sum_{s<s'}\int_{(\nu-1)\tau/2}^{\nu\tau/2}J_{ss'}^{(\nu)}(t)X^{(\nu)}_{ss'}(t) \mathrm{d}t.\label{q:EPR_deterministic}
\end{eqnarray}
The first law of thermodynamics then links the time-averaged power to the exchanged heat, i.e., $-{\overline {\cal P}} +{\overline Q_1}+{\overline Q_2}=0$. The entropy production can be written in terms of the heat as ${\overline \sigma}=-\sum_\nu \beta_\nu {\overline Q_\nu}$. 
Similarly, the efficiency is evaluated through $\eta= {\overline{\cal P}}/ {\overline Q_2 }$.
While the above expressions are exact for all-to-all interactions, they generally do not admit closed-form analytical solutions. We therefore proceed with a numerical approach.

{\it Universal  splitting of phase transitions}---Our first main results are summarized in Fig.~\ref{fig:fig1}(a) and Fig.~\ref{fig:fig1}(c), which show the order parameter $m$ as a function of the interaction energy $\epsilon$ for various switching rates $d$ (stochastic switching) or periods $\tau$ (deterministic switching), respectively, under a specific set of parameters.

We first note that all critical interaction energies, denoted here as $\epsilon_c^{A,B}$, are strongly influenced by the switching protocol and deviate significantly from those observed in the simultaneous case. Despite these differences, the qualitative classification of phase transitions remains unchanged: phase $A$ consistently undergoes a continuous phase transition at $\epsilon_c^A$, while phase $B$ exhibits a discontinuous transition at $\epsilon_c^B$.

\begin{figure}[htp]
    \includegraphics[width=\linewidth]{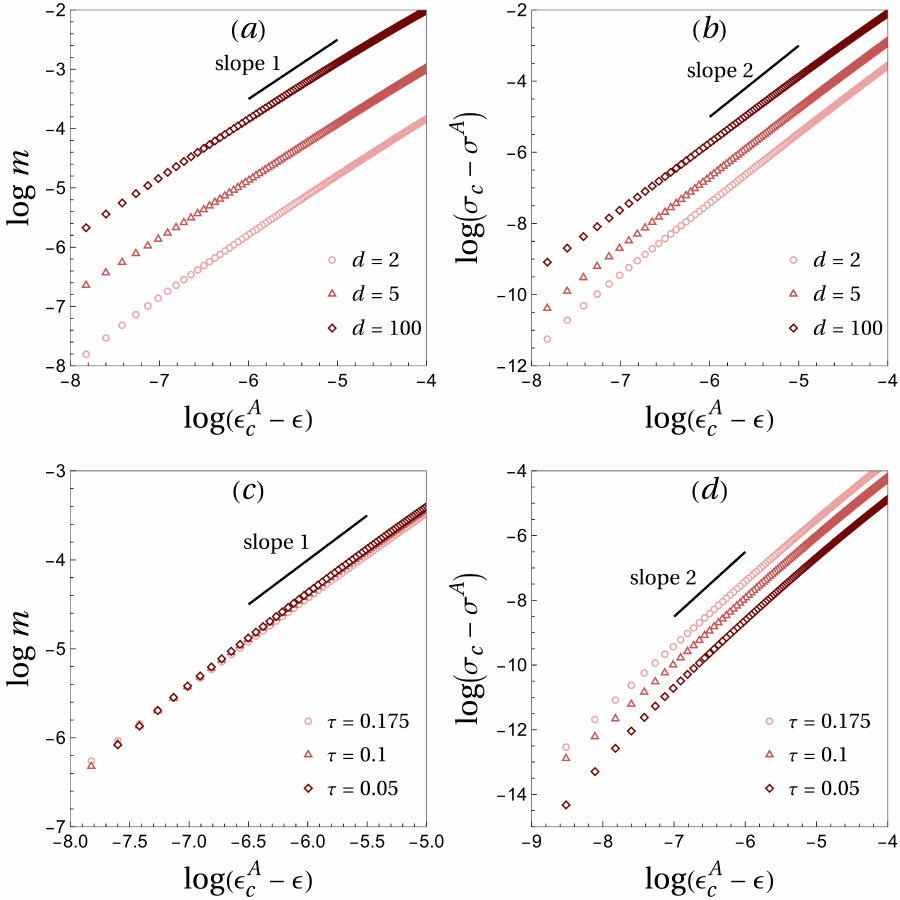}
     \caption{Critical scaling of the order parameter and entropy production near the critical point $\epsilon^{A}_c$. Panels (a) and (c) show the order parameter $m$ versus $\epsilon_c^{A}-\epsilon$; panels (b) and (d) display the reduced entropy production $\sigma_c-\sigma^A$ as a function of $\epsilon_c^{A}-\epsilon$. The top panels (a, b) correspond to the stochastic switching case for several values of the switching rate $d$, whereas the bottom panels (c d) correspond to the deterministic case for several cycle times $\tau$. Black solid lines are guides to the eye indicating power-law behavior, with slope $\beta=1$ for the order parameter and $2\beta$ for the entropy production.}
     \label{fig1c}
\end{figure}

For the stochastic case in particular, all critical interaction energies obey the inequality $\epsilon_c^B>\epsilon_c^A$. The $d$-dependence of the critical values $\epsilon_c^{A,B}$ is highly non-trivial but appears to scale as $d^{-1}$ for $d\gg1$, see Fig.~\ref{fig:fig1}(b). The behavior of transition points $\epsilon_c^{A,B}$ in the deterministic case is markedly different from that in the stochastic case in two key aspects. First, their dependence on the cycle period $\tau$ is strictly monotonic: increasing $\tau$ shifts both $\epsilon_c^{A}$ and $\epsilon_c^{B}$ toward lower values. In contrast, the stochastic case exhibits a non-monotonic dependence on $d^{-1}$ (which is qualitatively analogous to increasing $\tau$): initially, $\epsilon_c^{A,B}$ decreases with increasing $d^{-1}$, but beyond a certain point ($d^*\approx1/2$ and $1/4$ for phases A and B), this trend reverses and the critical energies begin to increase. Second, the relative ordering of the critical energies changes in a manner not observed for stochastic switching: for small $\tau$, one finds $\epsilon_c^{B} > \epsilon_c^{A}$, while at larger $\tau$, the inequality reverses. This crossover indicates the existence of an intermediate cycle period $\tau^* \approx 2.7$ at which the two critical lines intersect and both phase transitions occur at the same value of $\epsilon$.

Despite these differences, both switching protocols consistently show that the splitting of the phase transition into two distinct critical points is a robust feature arising from the interplay between temperature differences and non-conservative driving. Furthermore, in both cases all transition points converge to those of the simultaneous limit; in the stochastic protocol this convergence occurs as $d \to \infty$, while in the deterministic protocol, it is recovered in the fast-driving regime $\tau \to0$. In these limits, both descriptions approach the critical point $\epsilon_c^{A}$ given by Eq.~(\ref{eq2}). 

We now turn to the critical behavior of the order parameter and thermodynamic quantities, here exemplified by the entropy production $\sigma^{(A)}$. The presence of non-conservative driving and alternating coupling also modifies the critical behavior compared to other systems exhibiting $\mathbb{Z}_2$ symmetry~\cite{foraohzl3-hjnl}. This effect is illustrated in Fig.~\ref{fig1c}(a,~c) for both stochastic and deterministic switching protocols, where the order parameter scaling $m \sim (\epsilon_c^{A}-\epsilon)$ is consistent with a critical exponent $\beta=1$. To understand the behavior of entropy production near criticality, it is convenient to expand it in terms of the order parameter $m$ as $\sigma^{(A)} \sim \sigma_c + c_{\sigma} m^2 + \cdots$, where the expansion coefficients can be evaluated numerically. From the linear scaling for $m$, a quadratic scaling of the entropy production follows, i.e., $ \sigma^{(A)} - \sigma_c \sim (\epsilon^A_c - \epsilon)^2$, in contrast with the linear behavior for $F=0$ and/or $\beta_1 = \beta_2$ typically observed in standard $\mathbb{Z}_2$ symmetry-breaking transitions~\cite{noa2019,mamede2025collective,fiorefjtf-5glr}. Discontinuous phase transitions at $\epsilon^B_c$ are also manifested by a jump in the behavior of $\sigma$ (not shown).
\begin{figure*}[htp]
    \centering
    \begin{subfigure}{0.485\linewidth}
        \includegraphics[width=0.95\linewidth]{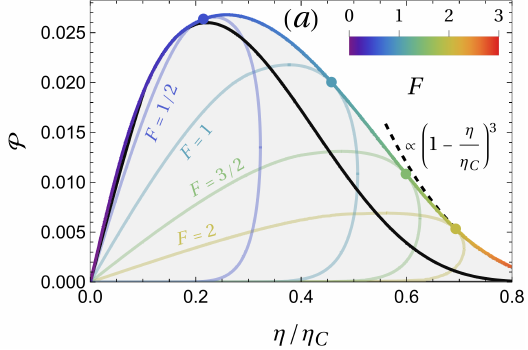}
    \end{subfigure}
    \begin{subfigure}{0.485\linewidth}
        \includegraphics[width=0.97\linewidth]{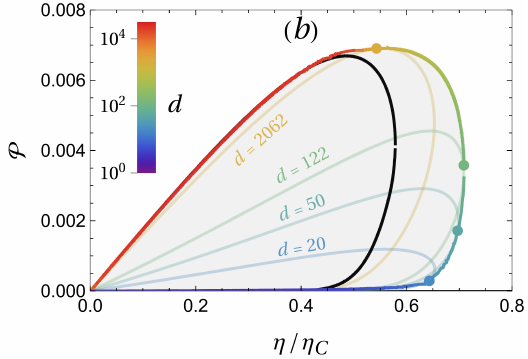}
    \end{subfigure}
    \caption{Power–efficiency Pareto fronts (rainbow–gradient lines) for \emph{stochastic} switching. {\bf(a)} $\epsilon,\,d$, and $F$ are varied simultaneously; the color gradient indicates how the optimal non-conservative drive $F$ evolves along the front. The dashed black line corresponds to $\mathcal{P}\sim(1-\eta/\eta_C)^{\alpha}$ with $\alpha=3$. {\bf(b)} Pareto fronts at fixed $F=2$, obtained by varying only $\epsilon$ and $d$. Here the gradient encodes the optimal $d$, demonstrating that finite-time switching yields globally superior performance---both in power and in efficiency---relative to simultaneous contact. In both panels, the black curves show that the simultaneous-contact limit $d\to\infty$ is entirely dominated by finite-time operation: operating the engine at (optimal) finite time always outperforms the simultaneous-contact limit. Colored lines in the suboptimal regions (shaded gray) represent, respectively, trade-offs at fixed $F$ (panel a) and fixed $d$ (panel b). Colored points mark where these fixed-parameter trade-offs become tangent to the corresponding Pareto fronts. Parameters are $\beta_2=1$ and $\beta_1=2$.
    }
    \label{fig:tradeoff_stochastic}
\end{figure*}

\begin{figure*}[htp]
    \centering
    \begin{subfigure}{0.485\linewidth}
        \includegraphics[width=0.95\linewidth]{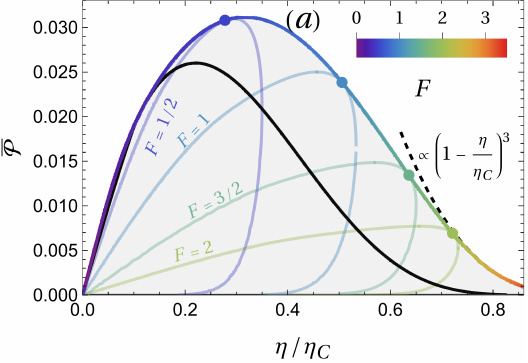}
    \end{subfigure}
    \begin{subfigure}{0.485\linewidth}
        \includegraphics[width=0.97\linewidth]{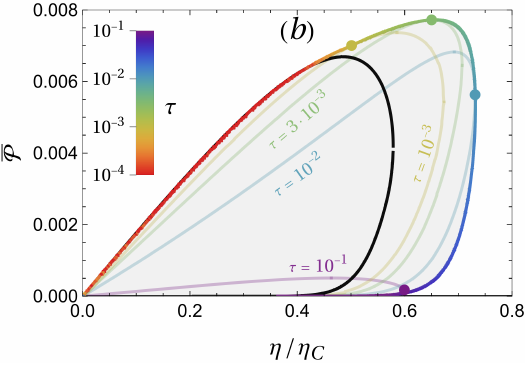}
    \end{subfigure}
    \caption{Power–efficiency Pareto fronts (rainbow–gradient lines) for \emph{deterministic} switching. {\bf(a)} $\epsilon,\,\tau$, and $F$ are varied simultaneously; the color gradient indicates how the optimal non-conservative drive $F$ evolves along the front. The dashed black line corresponds to $\overline{\mathcal{P}}\sim(1-\eta/\eta_C)^{\alpha}$ with $\alpha=3$.{\bf(b)} Pareto fronts at fixed $F=2$, obtained by varying only $\epsilon$ and $\tau$. Here the gradient encodes the optimal $\tau$, demonstrating that finite-time switching yields globally superior performance---both in power and in efficiency---relative to simultaneous contact. In both panels, the black curves show that the simultaneous-contact limit $\tau\to0$ is entirely dominated by finite-time operation: operating the engine at (optimal) finite time always outperforms the simultaneous-contact limit. Colored lines in the suboptimal regions (shaded gray) represent, respectively, trade-offs at fixed $F$ (panel a) and fixed $\tau$ (panel b). Colored points mark where these fixed-parameter trade-offs become tangent to the corresponding Pareto fronts. Parameters are $\beta_2=1$ and $\beta_1=2$.
    }
    \label{fig:tradeoff_deterministic}
\end{figure*}



{\it When to switch: optimal timing, Pareto trade-offs and their non-universal scaling}---We now address our second main goal: analyzing system performance beyond the simultaneous coupling scenario. For that, we turn to multi-objective optimization of the power and efficiency. The resulting trade-off relation between these quantities describes the full set of equally optimal solutions, including as particular cases the points of maximal efficiency and power. The curve that emerges is the Pareto front, which limits solution regions that are unattainable on one side of the front and suboptimal on the other side~\cite{forao2025characterization}. By tracing the front, one can find the set of underlying optimal parameters corresponding to different optimal points, e.g., for the maximal power or maximal efficiency. Note that within \emph{any} Pareto optimization setting, constraining one parameter produces a lower-dimensional Pareto problem whose solutions necessarily lie in the suboptimal region relative to the original, unconstrained optimization.
In our work, we compute the power-efficiency Pareto fronts using the NSGA-II genetic algorithm~\cite{pareto} at fixed reservoir temperatures $\beta_1,\,\beta_2$; remaining system parameters $\epsilon,\,F$ and $d$ (or $\tau$ in the deterministic case) are free to vary.

Figs.~\ref{fig:tradeoff_stochastic}(a) and~\ref{fig:tradeoff_deterministic}(a) show the full Pareto optimization landscape of power and efficiency for the stochastic and deterministic case, respectively. In both figures, rainbow-gradient lines represent the complete Pareto fronts obtained by optimizing simultaneously over $\epsilon$, $F$, and $d$ (or $\tau$), whereas gray-shaded regions contain suboptimal curves (fixed colors) corresponding to fixed-$F$ optimizations, where only $\epsilon$ and $d$ (or $\tau$) are varied. Each of these suboptimal curves becomes tangent to the global bounding Pareto front precisely at the point where its optimal values for $\epsilon$ and $d$ (or $\tau$) match those of the fully optimized front. The simultaneous-coupling limit, indicated by black curves, is fully encompassed by the global Pareto front; since fixing either $d\to\infty$ or $\tau=0$ constitutes a constraint on the optimization, the resulting Pareto fronts are necessarily suboptimal.
It is straightforward to see from the bounding Pareto fronts that for driving forces $F\leq~F_{\rm MP}~\approx~0.57$, i.e., at the point of maximal power, increasing $F$ first leads to an increase in \emph{both} the maximum efficiency and maximum power, given that $\epsilon$ and $d$ (or $\tau$) are allowed to freely assume their respective optimal values. This holds for both the stochastic and deterministic cases. For $F>F_{\rm MP}$, in contrast, increasing $F$ leads to a faster increase in maximum efficiency but drastically \emph{decreases} the maximum power that can be extracted. The Pareto front smoothly approaches zero as $F\to\infty$, and the suboptimal fronts with fixed $F$ collapse identically in this limit. 

Next, in Figs.~\ref{fig:tradeoff_stochastic}(b) and~\ref{fig:tradeoff_deterministic}(b), we show the constrained global Pareto fronts for a fixed non-equilibrium drive $F=2$, corresponding to the light-green suboptimal fronts in panels (a). The rainbow coloring now shows the value of the optimal rate $d$ or period $\tau$ along the front. Although the front shape differs substantially from the globally optimal curves in panels (a), the simultaneous optimization of power and efficiency in this case nonetheless produces a non-trivial Pareto front, since $\epsilon$ and $d$ (or $\tau$) are still allowed to vary simultaneously. 
Lines of fixed color in the suboptimal region (shaded gray) of panels (b) now indicate trade-offs where $d$ or $\tau$ are constrained to assume fixed values, in addition to a fixed $F$. Similar to panels (a), these curves become tangent to the unconstrained front in a single point. The simultaneous coupling limit (black curves) is then once again suboptimal, reaffirming that finite-time operation is globally superior. Since earlier results demonstrated independent optimizations of power and efficiency~\cite{filho2023powerful,foraohzl3-hjnl} where only $\epsilon$ was varied, it is instructive to analyze this case in more detail. Although the power is strongly suppressed for small $d$ (large $\tau$), corresponding to sparse switching in the stochastic case or long residence times in the deterministic case, both protocols display well-defined optima in the rainbow-colored front in panels (b). There exist unique values $d_{ME}$ ($\tau_{ME}$) that globally maximize the efficiency, as illustrated by the green curve in Fig.~\ref{fig:tradeoff_stochastic}(b) and the blue curve in Fig.~\ref{fig:tradeoff_deterministic}(b). Likewise, distinct values $d_{MP}$ ($\tau_{MP}$) globally maximize the power, shown by the yellow and green curves in Figs.~\ref{fig:tradeoff_stochastic}(b) and \ref{fig:tradeoff_deterministic}(b) for the stochastic and deterministic protocols, respectively.
Despite that for fixed $F$ and $d$ (or $\tau$) the resulting curve is, in a strict sense, a Pareto front, it is structurally trivial: its geometry is inherited entirely from the one-dimensional map $\epsilon\mapsto(\eta(\epsilon),\mathcal{P}(\epsilon))$ and there are no forbidden or dominated solutions within this set. Consequently, the front does not exhibit any \emph{genuine} multi-objective geometry; there is no underlying optimization involved. Similar arguments hold for fixing either $(\epsilon,F)$ or $(\epsilon,\tau)$ while varying $\tau$ or $F$, respectively.

\begin{figure}[htp]
    \centering
    \includegraphics[width=\linewidth]{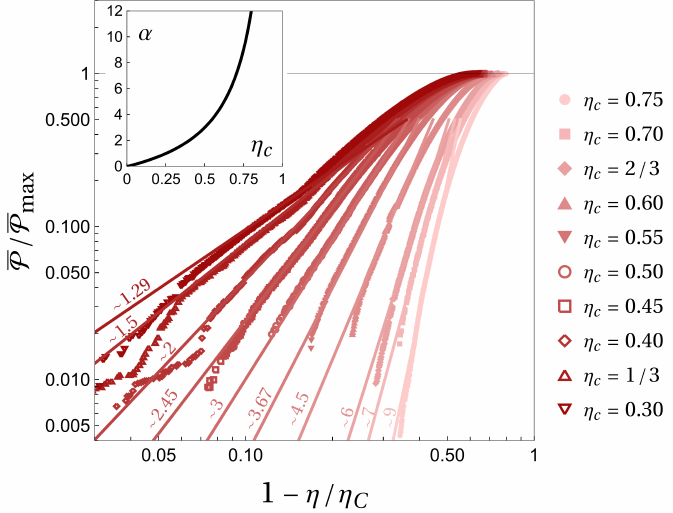}
    \caption{Rescaled power–efficiency Pareto fronts (colored symbols) for the deterministic switching protocol, shown on log–log axes for different reservoir temperature ratios, which fix the corresponding Carnot efficiency $\eta_C = 1-\beta_2/\beta_1$. As $\eta$ approaches $\eta_C$, the power follows a power-law decay $\overline{\mathcal{P}}\sim(1-\eta/\eta_C)^\alpha$, where $\alpha = 3\eta_C/(1-\eta_C)$  $\eta_C>1/4$ (see inset). Solid lines indicate power-law guides for the eye using this predicted exponent.}
    \label{fig:scaling_deterministic}
\end{figure}

Lastly, it is worth highlighting a remarkable feature about the global trade-off between power and efficiency as $\eta$ approaches $\eta_C$. In this case, by varying $\epsilon,\,\tau$ and $F$, power and efficiency are related via the power-law scaling $\overline{\mathcal{P}}\propto (1-\eta/\eta_C)^\alpha$, where the exponent $\alpha = 3\eta_C/(1-\eta_C)$ solely depends on the ratio between the reservoir temperatures. Despite the absence of a rigorous proof of this result, its validity has been verified empirically in both deterministic and stochastic cases; we show this in Fig.~\ref{fig:scaling_deterministic}  
for the deterministic case, for different choices of $\beta_2/\beta_1$ (see e.g. the inset for the relation between $\alpha$ and $\eta_C$). For $\eta_C < 1/4$ ---where the scaling becomes sublinear---the Pareto optimization scheme becomes numerically unstable and the estimation of $\alpha$ becomes imprecise. Although a similar relationship has been verified to hold for the simultaneous case, the exponent is different from the previous one.


Finally, we present a physical argument for why the maximum power achieved under the sequential protocol exceeds that of the simultaneous regime, in addition to the argument derived from Pareto optimality. In the latter setup, the system remains continuously coupled to both reservoirs, which inevitably produces a heat leak from hot bath to cold bath \cite{heatleak1,heatleak2}. This leakage bypasses the work-conversion mechanism: a portion of the energy injected at \(\beta_2\) flows directly to \(\beta_1\), irrespective of the system’s state, thereby reducing the amount of energy available to be transformed into useful work and ultimately suppressing the achievable power. Conversely, the contact with the reservoirs in the deterministic approach occurs in separate strokes, so no direct hot-to-cold transfer is possible within a single stage (the switching between thermal baths is a adiabatic process). As a result, each amount of heat absorbed from \(\beta_2\) is either converted into work during that stroke or stored in the system and later released to \(\beta_1\) in a controlled manner. The absence of instantaneous leakage can therefore ensure that a larger fraction of the injected energy remains available for work extraction, leading to a more favorable energy balance per cycle.

{\it Conclusions}---Our results reveal that the splitting of phase transitions is a feature that survives under any form of reservoir contact, including stochastic and deterministic switching protocols. This robustness demonstrates that the coexistence of two distinct ordered phases with fundamentally different critical and thermodynamical behavior is not a peculiarity of a specific setup, but a hallmark of collectively driven non-equilibrium systems.

We uncovered how temporal structure in the thermal contact reshapes the thermodynamic landscape. Both stochastic and deterministic coupling protocols exhibit optimal performance at intermediate switching rates or periods, where dissipation, ordering dynamics and cycle duration balance in a highly non-trivial manner. In these regimes, engines can outperform the simultaneous-contact limit in \emph{both} power and efficiency. This shows not only that finite-time operation can act as a genuine resource rather than a compromise, but also that collective behavior, non-equilibrium driving, and structured thermal contact can be deliberately exploited to enhance heat-engine performance through intrinsically non-equilibrium mechanisms.

In summary, this work advances the understanding of genuine non-equilibrium collective effects and their consequences for heat-engine performance. As promising directions for future research, we envision investigating other complex many-body settings (e.g., systems with energetic frustration), and study the role of asymmetric stochastic switching between the hot and cold reservoirs. Allowing different switching rates between baths will break time-reversal symmetry more strongly and may reveal richer phase structure, new performance regimes, or additional ways to control dissipation and fluctuations.




{\it Acknowledgments}---We acknowledge the financial support
from Brazilian agencies CNPq and FAPESP under grants No. 2021/03372-2, No. 2022/16192-5, No. 2022/15453-0, No. 2023/12490-4, 146224/2021-3 and No. 2024/03763-0.  {T.V.V.} was supported by JSPS KAKENHI Grant No.~JP23K13032. J.B. is supported by the Novo Nordisk Foundation with grant No. NNF18SA0035142.

\bibliography{refs}

@article{PhysRevE.110.054125,
  title = {Thermodynamics of underdamped Brownian collisional engines: General features and resonant phenomena},
  author = {For\~ao, Gustavo A. L. and Filho, Fernando S. and Akasaki, Bruno A. N. and Fiore, Carlos E.},
  journal = {Phys. Rev. E},
  volume = {110},
  issue = {5},
  pages = {054125},
  numpages = {10},
  year = {2024},
  month = {Nov},
  publisher = {American Physical Society},
  doi = {10.1103/PhysRevE.110.054125},
  url = {https://link.aps.org/doi/10.1103/PhysRevE.110.054125}
}

@article{heatleak1,
title = {Fundamental aspects of steady-state conversion of heat to work at the nanoscale},
journal = {Phys. Rep.},
volume = {694},
pages = {1-124},
year = {2017},
issn = {0370-1573},
doi = {https://doi.org/10.1016/j.physrep.2017.05.008},
url = {https://www.sciencedirect.com/science/article/pii/S0370157317301540},
author = {Giuliano Benenti and Giulio Casati and Keiji Saito and Robert S. Whitney}
}

@article{heatleak2,
   author = "Kosloff, Ronnie and Levy, Amikam",
   title = "Quantum Heat Engines and Refrigerators: Continuous Devices", 
   journal= "Annu. Rev. Phys. Chem.",
   year = "2014",
   volume = "65",
   number = "Volume 65, 2014",
   pages = "365-393",
   doi = "https://doi.org/10.1146/annurev-physchem-040513-103724",
   url = "https://www.annualreviews.org/content/journals/10.1146/annurev-physchem-040513-103724",
   publisher = "Annual Reviews",
   issn = "1545-1593",
   type = "Journal Article",
  }

@article{PhysRevLett.131.017102,
  title = {Tricritical Behavior in Dynamical Phase Transitions},
  author = {Agranov, Tal and Cates, Michael E. and Jack, Robert L.},
  journal = {Phys. Rev. Lett.},
  volume = {131},
  issue = {1},
  pages = {017102},
  numpages = {7},
  year = {2023},
  month = {Jul},
  publisher = {American Physical Society},
  doi = {10.1103/PhysRevLett.131.017102},
  url = {https://link.aps.org/doi/10.1103/PhysRevLett.131.017102}
}

@article{felipe,
  title = {Thermodynamics of a minimal interacting heat engine: Comparison between engine designs},
  author = {Hawthorne, Felipe and Cleuren, B. and Fiore, Carlos E.},
  journal = {Phys. Rev. E},
  volume = {109},
  issue = {6},
  pages = {064120},
  numpages = {12},
  year = {2024},
  month = {Jun},
  publisher = {American Physical Society},
  doi = {10.1103/PhysRevE.109.064120},
  url = {https://link.aps.org/doi/10.1103/PhysRevE.109.064120}
}

@article{danielPhysRevResearch.2.043257,
  title = {Coarse-grained entropy production with multiple reservoirs: Unraveling the role of time scales and detailed balance in biology-inspired systems},
  author = {Busiello, Daniel M. and Gupta, Deepak and Maritan, Amos},
  journal = {Phys. Rev. Res.},
  volume = {2},
  issue = {4},
  pages = {043257},
  numpages = {14},
  year = {2020},
  month = {Nov},
  publisher = {American Physical Society},
  doi = {10.1103/PhysRevResearch.2.043257},
  url = {https://link.aps.org/doi/10.1103/PhysRevResearch.2.043257}
}

@article{friedemann2018quantum,
  title={Quantum tricritical points in NbFe2},
  author={Friedemann, Sven and Duncan, Will J and Hirschberger, Max and Bauer, Thomas W and K{\"u}chler, Robert and Neubauer, Andreas and Brando, Manuel and Pfleiderer, Christian and Grosche, F Malte},
  journal={Nat. Phys.},
  volume={14},
  number={1},
  pages={62--67},
  doi={10.1038/nphys4242},
  year={2018},
  publisher={Nature Publishing Group UK London}
}

@article{mamede2025collective,
  title={Collective heat engines via different interactions: Minimal models, thermodynamics and phase transitions},
  author={Mamede, Iago N and Henkes, Vit{\'o}ria T and Fiore, Carlos E},
  journal={arXiv preprint arXiv:2508.06438},
  year={2025}
}

@Article{busiello2021dissipation,
author={Busiello, Daniel Maria
and Liang, Shiling
and Piazza, Francesco
and De Los Rios, Paolo},
title={Dissipation-driven selection of states in non-equilibrium chemical networks},
journal={Commun. Chem.},
year={2021},
volume={4},
number={1},
pages={16},
doi={10.1038/s42004-021-00454-w},
url={https://doi.org/10.1038/s42004-021-00454-w}
}

@article{Meibohm24,
  title = {Minimum-dissipation principle for synchronized stochastic oscillators far from equilibrium},
  author = {Meibohm, Jan and Esposito, Massimiliano},
  journal = {Phys. Rev. E},
  volume = {110},
  issue = {4},
  pages = {L042102},
  numpages = {6},
  year = {2024},
  month = {Oct},
  publisher = {American Physical Society},
  doi = {10.1103/PhysRevE.110.L042102},
  url = {https://link.aps.org/doi/10.1103/PhysRevE.110.L042102}
}

@article{gatien2,
  title = {Efficiency Fluctuations of Stochastic Machines Undergoing a Phase Transition},
  author = {Vroylandt, Hadrien and Esposito, Massimiliano and Verley, Gatien},
  journal = {Phys. Rev. Lett.},
  volume = {124},
  issue = {25},
  pages = {250603},
  numpages = {6},
  year = {2020},
  month = {Jun},
  publisher = {American Physical Society},
  doi = {10.1103/PhysRevLett.124.250603},
  url = {https://link.aps.org/doi/10.1103/PhysRevLett.124.250603}
}

@article{rolandi,
  title = {Collective Advantages in Finite-Time Thermodynamics},
  author = {Rolandi, Alberto and Abiuso, Paolo and Perarnau-Llobet, Mart\'{\i}},
  journal = {Phys. Rev. Lett.},
  volume = {131},
  issue = {21},
  pages = {210401},
  numpages = {7},
  year = {2023},
  month = {Nov},
  publisher = {American Physical Society},
  doi = {10.1103/PhysRevLett.131.210401},
  url = {https://link.aps.org/doi/10.1103/PhysRevLett.131.210401}
}

@article{lynn2021broken,
author = {Christopher W. Lynn  and Eli J. Cornblath  and Lia Papadopoulos  and Maxwell A. Bertolero  and Danielle S. Bassett },
title = {Broken detailed balance and entropy production in the human brain},
journal = {PNAS},
volume = {118},
number = {47},
pages = {e2109889118},
year = {2021},
doi = {10.1073/pnas.2109889118},
URL = {https://www.pnas.org/doi/abs/10.1073/pnas.2109889118}
}

@article{gatien,
	doi = {10.1209/0295-5075/120/30009},
	url = {https://doi.org/10.1209/0295-5075/120/30009},
	year = 2017,
	month = {nov},
	publisher = {{IOP} Publishing},
	volume = {120},
	number = {3},
	pages = {30009},
	author = {Hadrien Vroylandt and Massimiliano Esposito and Gatien Verley},
	title = {Collective effects enhancing power and efficiency},
	journal = {{EPL}}
}

@article{herpich2,
  title = {Universality in driven Potts models},
  author = {Herpich, Tim and Esposito, Massimiliano},
  journal = {Phys. Rev. E},
  volume = {99},
  issue = {2},
  pages = {022135},
  numpages = {7},
  year = {2019},
  month = {Feb},
  publisher = {American Physical Society},
  doi = {10.1103/PhysRevE.99.022135},
  url = {https://link.aps.org/doi/10.1103/PhysRevE.99.022135}
}

@article{herpich,
  title = {Collective Power: Minimal Model for Thermodynamics of Nonequilibrium Phase Transitions},
  author = {Herpich, Tim and Thingna, Juzar and Esposito, Massimiliano},
  journal = {Phys. Rev. X},
  volume = {8},
  issue = {3},
  pages = {031056},
  numpages = {20},
  year = {2018},
  month = {Sep},
  publisher = {American Physical Society},
  doi = {10.1103/PhysRevX.8.031056},
  url = {https://link.aps.org/doi/10.1103/PhysRevX.8.031056}
}

@article{rosas1,
  title = {Stochastic thermodynamics for a periodically driven single-particle pump},
  author = {Rosas, Alexandre and Van den Broeck, Christian and Lindenberg, Katja},
  journal = {Phys. Rev. E},
  volume = {96},
  issue = {5},
  pages = {052135},
  numpages = {7},
  year = {2017},
  month = {Nov},
  publisher = {American Physical Society},
  doi = {10.1103/PhysRevE.96.052135},
  url = {https://link.aps.org/doi/10.1103/PhysRevE.96.052135}
}

@article{rosas2,
  title = {Three-stage stochastic pump: Another type of Onsager-Casimir symmetry and results far from equilibrium},
  author = {Rosas, Alexandre and Van den Broeck, Christian and Lindenberg, Katja},
  journal = {Phys. Rev. E},
  volume = {97},
  issue = {6},
  pages = {062103},
  numpages = {7},
  year = {2018},
  month = {Jun},
  publisher = {American Physical Society},
  doi = {10.1103/PhysRevE.97.062103},
  url = {https://link.aps.org/doi/10.1103/PhysRevE.97.062103}
}

@article{mamede2021obtaining,
  title = {Obtaining efficient thermal engines from interacting Brownian particles under time-periodic drivings},
  author = {Mamede, Iago N. and Harunari, Pedro E. and Akasaki, Bruno A. N. and Proesmans, Karel and Fiore, C. E.},
  journal = {Phys. Rev. E},
  volume = {105},
  issue = {2},
  pages = {024106},
  numpages = {11},
  year = {2022},
  month = {Feb},
  publisher = {American Physical Society},
  doi = {10.1103/PhysRevE.105.024106},
  url = {https://link.aps.org/doi/10.1103/PhysRevE.105.024106}
}

@article{noa2021efficient,
  title = {Efficient asymmetric collisional Brownian particle engines},
  author = {Noa, C. E. Fern\'andez and Stable, Angel L. L. and Oropesa, William G. C. and Rosas, Alexandre and Fiore, C. E.},
  journal = {Phys. Rev. Res.},
  volume = {3},
  issue = {4},
  pages = {043152},
  numpages = {13},
  year = {2021},
  month = {Dec},
  publisher = {American Physical Society},
  doi = {10.1103/PhysRevResearch.3.043152},
  url = {https://link.aps.org/doi/10.1103/PhysRevResearch.3.043152}
}

@article{foraohzl3-hjnl,
  title = {Splitting of nonequilibrium phase transitions in driven Ising models},
  author = {For\~ao, Gustavo A. L. and Filho, Fernando S. and Vieira, Andr\'e P. and Cleuren, Bart and Busiello, Daniel M. and Fiore, Carlos E.},
  journal = {Phys. Rev. Res.},
  volume = {7},
  issue = {3},
  pages = {L032049},
  numpages = {7},
  year = {2025},
  month = {Sep},
  publisher = {American Physical Society},
  doi = {10.1103/hzl3-hjnl},
  url = {https://link.aps.org/doi/10.1103/hzl3-hjnl}
}

@article{mamede2023,
  title = {Thermodynamics of interacting systems: The role of the topology and collective effects},
  author = {Mamede, Iago N. and Proesmans, Karel and Fiore, Carlos E.},
  journal = {Phys. Rev. Res.},
  volume = {5},
  issue = {4},
  pages = {043278},
  numpages = {12},
  year = {2023},
  month = {Dec},
  publisher = {American Physical Society},
  doi = {10.1103/PhysRevResearch.5.043278},
  url = {https://link.aps.org/doi/10.1103/PhysRevResearch.5.043278}
}

@article{noa2019,
  title = {Entropy production as a tool for characterizing nonequilibrium phase transitions},
  author = {Noa, C. E. Fern\'andez and Harunari, Pedro E. and de Oliveira, M. J. and Fiore, C. E.},
  journal = {Phys. Rev. E},
  volume = {100},
  issue = {1},
  pages = {012104},
  numpages = {10},
  year = {2019},
  month = {Jul},
  publisher = {American Physical Society},
  doi = {10.1103/PhysRevE.100.012104},
  url = {https://link.aps.org/doi/10.1103/PhysRevE.100.012104}
}

@article{noa2020thermodynamics,
  title = {Thermodynamics of collisional models for Brownian particles: General properties and efficiency},
  author = {Stable, Angel L. L. and Noa, C. E. Fern\'andez and Oropesa, William G. C. and Fiore, C. E.},
  journal = {Phys. Rev. Res.},
  volume = {2},
  issue = {4},
  pages = {043016},
  numpages = {9},
  year = {2020},
  month = {Oct},
  publisher = {American Physical Society},
  doi = {10.1103/PhysRevResearch.2.043016},
  url = {https://link.aps.org/doi/10.1103/PhysRevResearch.2.043016}
}

@article{harunari2020maximal,
  title = {Maximal power for heat engines: Role of asymmetric interaction times},
  author = {Harunari, Pedro E. and Filho, Fernando S. and Fiore, Carlos E. and Rosas, Alexandre},
  journal = {Phys. Rev. Res.},
  volume = {3},
  issue = {2},
  pages = {023194},
  numpages = {7},
  year = {2021},
  month = {Jun},
  publisher = {American Physical Society},
  doi = {10.1103/PhysRevResearch.3.023194},
  url = {https://link.aps.org/doi/10.1103/PhysRevResearch.3.023194}
}

@article{fiorefjtf-5glr,
  title = {Emergent collective heat engines from neighborhood-dependent thermal reservoirs},
  author = {Fiore, Carlos E.},
  journal = {Phys. Rev. E},
  volume = {112},
  issue = {6},
  pages = {064105},
  numpages = {6},
  year = {2025},
  month = {Dec},
  publisher = {American Physical Society},
  doi = {10.1103/fjtf-5glr},
  url = {https://link.aps.org/doi/10.1103/fjtf-5glr}
}

@article{forao2025characterization,
doi = {10.1088/1367-2630/adf08d},
url = {https://doi.org/10.1088/1367-2630/adf08d},
year = {2025},
month = {jul},
publisher = {IOP Publishing},
volume = {27},
number = {7},
pages = {074605},
author = {Forão, Gustavo A L and Berx, Jonas and Fiore, Carlos E},
title = {Characterization and optimization of heat engines: Pareto-optimal fronts and universal features},
journal = {New J. Phys.}
}

@article{Imparato_2015,
doi = {10.1088/1367-2630/17/12/125004},
url = {https://dx.doi.org/10.1088/1367-2630/17/12/125004},
year = {2015},
month = {dec},
publisher = {IOP Publishing},
volume = {17},
number = {12},
pages = {125004},
author = {Alberto Imparato},
title = {Stochastic thermodynamics in many-particle systems},
journal = {New J. Phys.}
}

@article{filho2023powerful,
  title = {Powerful ordered collective heat engines},
  author = {Filho, Fernando S. and For\~ao, Gustavo A. L. and Busiello, Daniel M. and Cleuren, B. and Fiore, Carlos E.},
  journal = {Phys. Rev. Res.},
  volume = {5},
  issue = {4},
  pages = {043067},
  numpages = {12},
  year = {2023},
  month = {Oct},
  publisher = {American Physical Society},
  doi = {10.1103/PhysRevResearch.5.043067},
  url = {https://link.aps.org/doi/10.1103/PhysRevResearch.5.043067}
}

@BOOK{pareto,
  title     = "{Multi-Objective} Optimization using Evolutionary Algorithms",
  author    = "Deb, Kalyanmoy",
  publisher = "John Wiley \& Sons",
  series    = "Wiley Interscience Series in Systems and Optimization",
  year      =  2001,
    isbn = {047187339X}
}

@article{Hagman2025,
doi = {10.1088/1367-2630/ae18be},
url = {https://doi.org/10.1088/1367-2630/ae18be},
year = {2025},
publisher = {IOP Publishing},
volume = {27},
number = {11},
pages = {114507},
author = {Hagman, Rasmus and Berx, Jonas and Splettstoesser, Janine and Kirchberg, Henning},
title = {Optimising finite-time quantum information engines using Pareto bounds},
journal = {New J. Phys.}
}

@BOOK{Golubitsky1984,
  title     = "Singularities and Groups in Bifurcation Theory",
  author    = "Golubitsky, Martin and Schaeffer, David G",
  publisher = "Springer",
  series    = "Applied Mathematical Sciences",
  edition   =  1,
  year      =  1984,
  address   = "New York, NY",
    doi = "https://doi.org/10.1007/978-1-4612-5034-0"
}

@BOOK{Landau1996,
  title     = "Statistical Physics",
  author    = "Landau, L D and Lifshitz, E M",
  publisher = "Butterworth-Heinemann",
  edition   =  3,
  year      =  1996,
  address   = "Oxford, England"
}
\end{document}